# Signatures of Molecular Magnetism in Single-Molecule Transport Spectroscopy

Moon-Ho Jo,[1,5*] Jacob E. Grose,[2*] Kanhayalal Baheti,[3] Mandar M. Deshmukh,[1] Jennifer J. Sokol,[3] Evan M. Rumberger,[4] David N. Hendrickson,[4] Jeffrey R. Long,[3] Hongkun Park[1†] and D. C. Ralph[2†]

[1]*Department of Chemistry & Chemical Biology and Department of Physics, Harvard University, 12 Oxford Street, Cambridge, MA 02138, USA,* [2]*Department of Physics, Cornell University, Ithaca, NY 14853, USA,* [3]*Department of Chemistry, University of California, Berkeley, CA 94720, USA,* [4]*Department of Chemistry and Biochemistry, University of California, San Diego, CA 92093, USA,* [5]*Department of Materials Science and Engineering, POSTECH, Pohang, KyungBuk Do, Republic of Korea.*

[*]These authors contributed equally to this work.

[†]e-mail: ralph@ccmr.cornell.edu; Hongkun_Park@harvard.edu

**Single-molecule transistors provide a unique experimental tool to investigate the coupling between charge transport and the molecular degrees of freedom in individual molecules. One interesting class of molecules for such experiments are the single-molecule magnets, since the intramolecular exchange forces present in these molecules should couple strongly to the spin of transport electrons, thereby providing both new mechanisms for modulating electron flow and also new means for probing nanoscale magnetic excitations. Here we report single-molecule transistor measurements on devices incorporating $Mn_{12}$ molecules. By studying the electron-tunneling spectrum as a function of magnetic field, we are able to identify clear signatures of magnetic states and their associated magnetic anisotropy. A comparison of the data to simulations also suggests that electron flow can strongly enhance magnetic relaxation of the magnetic molecule.**

When electrons traverse a molecule, their flow can be affected by interactions with other degrees of freedom in the molecule[1-4]. Previous studies have shown that transport spectroscopy of these structures can provide information on discrete energy excitations associated with electronic[2-4] and vibrational[1,3,4] degrees of freedom. Here we report an extension of such measurements to a prototypical single-molecule magnet $Mn_{12}O_{12}(O_2C-R)_{16}(H_2O)_4$ depicted in Fig. 1(a) (henceforth "$Mn_{12}$"; here *R* represents a generic chemical functional group)[5]. We examine the manner in which collective spin states of the molecule couple to electron transport, and in particular how this affects the magnetic-field (*B*) dependence of the measured level spectra. We find that low-temperature current-voltage measurements reveal two signatures of magnetic states and magnetic anisotropy: energy splitting at zero $B$ within low-energy spin manifolds and a nonlinear evolution of energy level positions with $B$. The zero-field splitting varies from device to device, and we interpret this as evidence for magnetic anisotropy variations upon changes in molecular geometry and environment. We do not observe hysteresis in the electron-tunneling spectrum as a function of swept magnetic field, as one might expect to find in analogy to magnetization measurements on large ensembles of $Mn_{12}$ molecules in bulk crystals[6-9]. We ascribe the absence of hysteresis in the single-molecule transistors to magnetic relaxation caused by tunneling electrons, and we discuss this effect in comparison to a simple model calculation. We note that Heersche[10] *et al.* recently posted an independent study of single-$Mn_{12}$ transistors. They reported little about the magnetic-field dependence of the spectra, which is our primary focus.

Our devices are made by applying dilute (~100 $\mu$M) solutions of $Mn_{12}O_{12}(O_2CCH_3)_{16}(H_2O)_4$ ("$Mn_{12}Ac$") in acetonitrile or $Mn_{12}O_{12}(O_2CCHCl_2)_{16}(H_2O)_4$ ("$Mn_{12}Cl$") in methylene chloride to gold wires, and then creating molecular-sized gaps in the wires by electromigration[1-3]. An oxidized aluminum layer beneath the device serves as a gate electrode, and the measurements are conducted at temperatures $T \leq 300$ mK. The fabrication and measurement procedures are described in more detail in the Methods Section. In addition to the devices made with molecules, we conducted control experiments by preparing approximately 80 junctions from the same wafers using solvent alone. We observed simple linear tunneling *I-V* characteristics or no measurable current in all but two of the control samples. In those two, we found Coulomb blockade characteristics; however these devices exhibited small charging energies (< 75 meV) as compared to much larger charging energies (> 250 meV) in molecular devices, and could therefore be ascribed to nanoscale metal particles created by the electromigration process[11]. None of the control samples displayed any of the characteristics that we associate below with the existence of magnetic states (see below).

We fabricated more than 70 chips of devices incorporating either $Mn_{12}Ac$ or $Mn_{12}Cl$ molecules, each chip containing more than 40 tunneling junctions, although not every junction on a chip was measured. Approximately 10% of these 70 chips exhibited devices with Coulomb blockade characteristics with a yield of 1-4 devices per chip (the other devices exhibited either characteristics of simple tunnel junctions or no measurable current whatsoever). In total, 16 junctions exhibiting Coulomb blockade were sufficiently stable over time for thorough investigation.

It should be noted that our measurements do not provide a means to verify that the $Mn_{12}$ molecules remain intact through the processes of deposition and electromigration. It is possible that the molecules may lose water ligands, degrade into smaller magnetic subunits or aggregate into larger clusters. However, the qualitative conclusions that we present about tunneling via a magnetic molecule are unaffected regardless of whether our data reflect tunneling through an intact $Mn_{12}$ molecule or through a smaller or larger cluster with nonzero total spin.

Figure 1(b) shows *I-V* curves of a single-$Mn_{12}Ac$ transistor as a function of gate voltage ($V_g$). The current is suppressed up to a certain threshold *V*, which varies linearly with $V_g$. This is a signature of Coulomb blockade. Outside of the blockade region, *I* increases in a stepwise manner with increasing *V*, corresponding to tunneling via discrete quantum-mechanical states.

Figure 2 shows plots of the differential conductance ($dI/dV$) of single-$Mn_{12}$ transistors as a function of $V$ and $V_g$. In these plots we observe crossed diagonal lines intersecting at $V = 0$ that indicate tunneling transitions between the ground energy levels of adjacent charge states. In Fig. 2(a), two additional peaks in $dI/dV$ (marked with green and yellow arrows) can also be observed, corresponding to excited-state transitions with energies of ~1.1 meV and 1.34 meV. These excitations behave differently with the applied magnetic field $B$, as shown in Fig. 2(b). Specifically, as $B$ is raised to 8 T, the peak marked by the yellow arrow moves to higher energy, while the peak marked by green does not shift. This observation indicates that the peak marked by green arises from a non-magnetic origin, most likely from excited vibrational levels of a molecule[1,12,13].

Figures 2(c) and (d) show $dI/dV$ vs. $V$ and $V_g$ plots for a single-$Mn_{12}$Cl transistor at $B = 0$ and $B = 8$ T. Here we do not directly resolve separate ground and excited levels at $B = 0$, but instead observed inelastic cotunneling features within the left blockade region[14]. This cotunneling feature indicates the existence of an excited state with a small energy splitting (~ 0.25 meV) from the ground state. As Fig. 2(d) illustrates, this state moves to higher energy and becomes separately resolvable in the sequential-tunneling spectrum as $B$ increases.

In total, we have observed a zero-field splitting which then increases as a function of $B$ in four different devices. The magnitude of the zero-field splitting ranged from 0.25 meV to 1.34 meV (both extremes are represented in Fig. 2). The other twelve devices that we studied in detail also exhibited Zeeman splitting, but with apparent degeneracy at $B = 0$ T.

The presence of zero-field splitting is a signature of magnetism in a quantum system. In non-magnetic quantum-dot systems, the absence of any zero-field splitting for Zeeman-split levels is observed universally (as required by Kramers' Theorem) for tunneling transitions from even ($S = 0$) to odd ($S = 1/2$) numbers of electrons. For odd-to-even transitions, only one spin transition is allowed for the lowest-energy tunneling transition because of Pauli blocking, but degeneracies at $B = 0$ are still generally observed for excited states[15]. In quantum dots made from ferromagnetic nanoparticles, the presence of zero-field splitting has been observed previously[16,17] due to magnetic anisotropy that affects tunneling transitions between states with $S \geq 1/2$[16,18,19]. The presence of zero-field-split energy levels in four of our devices therefore demonstrates that tunneling in these devices is occurring via magnetic states with non-zero magnetic anisotropy.

In $Mn_{12}$, the magnetic anisotropy is associated with Jahn-Teller distortions in the octahedral coordination spheres of the eight $Mn^{3+}$ ions, and thus it is sensitive to changes in the charge state[20] and environment of the molecule[20-24]. For example, the molecule $Mn_{12}O_{12}(O_2CCF_3)_{16}(H_2O)_4$ has been isolated in two different crystal forms, for which differences in the Jahn-Teller distortion axes around the $Mn^{3+}$ ions give rise to different zero-field splitting parameters of $D = $ -0.042 and -0.081 meV[24]. We therefore suggest that the variations in zero-field splitting observed between devices are due to variations in the environment of each molecule as it interacts with the surface in the device, if not variations in the overall molecular structure.

Figure 3(a) illustrates the $B$ dependence of 1.34 meV excitation (seen in Fig. 2) for fixed $V_g = $ -1.768 V. Here $B$ was swept slowly from –8 T to 8 T while more rapid scans of $dI/dV$ vs. $V$ were measured. The variations of the transition energies are continuous, symmetric around $B = 0$, and show deviations from perfect linearity. We observe no hysteresis upon reversing the direction of the field sweep.

To analyze these observations, we consider a simple model for electron tunneling via an individual magnetic molecule. We start with a Hamiltonian that includes the lowest-order uniaxial magnetic anisotropy along the z-axis and the Zeeman energy corresponding to a magnetic field applied in an arbitrary direction:

$$H = D_N S_Z^2 + g\mu_B \mathbf{B}\cdot\mathbf{S}. \qquad (1)$$

Here, $D_N$ is the anisotropy constant for the multiplet with $N$ electrons, $S_Z$ is the z-component of $S$, $g$ is the electronic $g$-factor and $\mu_B$ is the Bohr magneton. The value of the $g$-factor is known to be close to 2.0 for bulk $Mn_{12}$ samples[5]. Electrochemical measurements have shown that the oxidation states $[Mn_{12}]^{1-}$, $[Mn_{12}]^0$ and $[Mn_{12}]^{1+}$ of the $Mn_{12}$ molecules are reversible[23], and from magnetic measurements of the $Mn_{12}Cl$ series, the ground state spin of $[Mn_{12}]^0$, $[Mn_{12}]^{1-}$ and $[Mn_{12}]^{2-}$ species were found to be $S = 10$, 19/2 and 10, respectively[24]. We assume that the tunneling energies we measure in Fig. 2(a) and (b) and in Fig. 3(a) correspond to transitions between an $S = 19/2$, ($N$-1)-electron state and an $S = 10$, $N$-electron state. Other choices of spin states produce qualitatively similar results, as long as the difference in total spin remains 1/2. A diagram of the energy levels for Eq. (1) with $S = 10$ and $B = 0$ is pictured in Fig. 3(b). For computational simplicity, we assumed that only states in the ground-state spin multiplet of each charge state are accessible in our experiment, although experiments suggest that the $S = 9$ manifold might lie several meV above the ground $S = 10$ state of neutral $Mn_{12}$[25].

From Eq. (1), we can identify that the magnitude of the zero field splitting observed at positive bias in Fig. 3(a) is due to the energy difference $D_N\Delta(S_Z^2)$ between the $S_z = -S$ ground state and the first excited state, $S_z = -S+1$, for the charge state with the higher spin $S$. From the observed energy gap of 1.34 meV in Fig. 2(a), we estimate $D_N \sim -0.071$ meV for the higher-spin charge state in this sample (assuming $S = 10$). Measurements on bulk crystals find $-0.057$ meV for the neutral $Mn_{12}$ species[26] and $-0.077$ meV for a singly reduced $Mn_{12}$ derivative[20].

For more detailed comparisons to the data, we simulated the current flow through a device by determining the energy eigenstates of a $Mn_{12}$ molecule through numerical diagonalization of Eq. (1) for both the $S = 19/2$ and $S = 10$ ground states and then by calculating the *I-V* curve using a master equation approach to account for energetically-allowed tunneling transitions. We included only lowest-order sequential tunneling processes. In our main simulation, we did not simply assume steady-state occupation probabilities (this assumption will necessarily eliminate hysteresis) because transitions out of metastable magnetic states may require times scales much longer than our experiment so that a true steady-state probability distribution may not be relevant experimentally. We instead integrated the master equation with respect to time at each incremented value of *V* and/or *B*, calculated the distribution of occupation probabilities corresponding to the experimental time scale, and then obtained *I* based on this distribution. We allowed occupation of states corresponding to a reversed magnetic moment only if the transition rate to these states is relevant experimentally. When questions of hysteresis were not at issue, we also performed the simpler alternative calculation that assumes steady-state occupation probabilities.

The results of our simulations, for parameter values chosen to mimic Fig. 3(a) (there are too many free parameters to claim quantitative fits), are shown in Fig. 4(a-c). The parameters for the main simulation in Fig. 4(a,b) are $S_{N-1} = 19/2$, $S_N = 10$, $D_N = D_{N-1} =$ −0.071 meV, g = 2.0, a field angle of 45 degrees with respect to the easy axis, capacitance ratios $C_{gate}:C_{source}:C_{drain}$ = 1:13:4 determined from the slopes of the tunneling thresholds in Fig. 2(a,b), bare tunneling rates $\Gamma_{source}$ = 8 GHz and $\Gamma_{drain}$ = 0.8 GHz, and a magnetic field sweep rate of 0.02 T/min. The simulation in Fig. 4(c) employs the same parameters except that we assume different anisotropy energies for the different charge states, $D_{N-1}$ = -0.071 meV and $D_N$ = -0.086 meV, and for Fig. 4(c) we employ the simpler quasi-steady state algorithm. These calculations capture many of the qualitative features of the data.

The nonlinear variation of the energy levels vs. B shown in Fig. 3(a) can be explained as a consequence of anisotropy in a magnetic molecule[16,17]. When $D_N = D_{N-1}$ (Fig. 4(a,b)) we find no deviation from linearity, but the sign of the curvature that is observed in Fig. 3(a) emerges naturally if the magnetic anisotropy parameter $D_N$ is larger for the higher-spin charge state than the smaller-spin state (Fig. 4(c)). A quantitative determination of $D_N$ for the smaller-spin charge state is difficult because the angle between B and the anisotropy axis is not known.

The simulation (Fig. 4(a,b)) also reproduces naturally the number of energy levels measured in the experiment: just two at positive V and one at negative V, even though we assume spins of S = 10 and 19/2. The reason is that just two levels are observed at positive V is that a voltage high enough to cause tunneling to the first excited state also permits tunneling transitions higher up the ladder of spin states because transitions to higher spin states all require progressively less energy. This is consistent with our observation of only one magnetic excited state in all four devices in which we found zero-field splittings. The observation of only ground state tunneling at negative $V$ can be explained as a consequence of an asymmetry in the coupling of the molecule to the two electrodes, which can cause excited states to contribute negligibly to the total current for one bias direction[27]. The excited magnetic level must be associated with the higher-spin charge state; this implies that the higher-spin state is associated with the more-positive-$V_g$ charge state in Fig. 2(a,b), but the more-negative-$V_g$ charge state in Fig. 2(c,d).

Finally, we comment on the absence of hysteresis in the electron-tunneling spectrum measured as a function of magnetic field. Magnetic measurements of macroscopic $Mn_{12}$ crystals find low-temperature hysteresis for all field directions except those within a few degrees of the hard magnetic axis[8,9]. As illustrated in Fig. 3(a), however, we do not observe hysteresis in any of our $Mn_{12}$ transistors within a magnetic field sweep range of ±8 T, at $T \leq 300$ mK. This lack of hysteresis is also in contrast to related measurements of energy-level spectra for magnetic Co nanoparticles, with higher spin S ≈ 1000[16,17].

One possible explanation for the absence of hysteresis in the tunneling spectra of the molecular magnets is that repeated spin transitions caused by tunneling electrons may allow the molecule to surmount the anisotropy barrier through the sequential occupation of increasingly high-energy spin states. This could allow the magnetic moment to undergo excursions between the $S_z < 0$ and $S_z > 0$ energy wells on time scales fast compared to the sweep rate of $B$, so that there are no abrupt changes measured in the tunneling spectrum as a function of $B$. For the case of magnetic nanoparticles, Waintal and Brouwer[28] predicted precisely this scenario - under certain conditions tunneling electrons can induce magnetic relaxation despite the presence of anisotropy. This conclusion also follows from our simulations based on Eq. (1). We find that if, during the magnetic-field sweeps, the voltage is scanned to values sufficiently high to measure the first excited state of the higher-spin state, then there is no hysteresis (Fig. 4(a)), in agreement with experiment. This voltage is sufficiently high to enable tunneling for all allowed tunneling transitions with $\Delta S = \pm 1/2$, so that all states in the spin multiplets are accessible despite the magnetic anisotropy. Within the approximation that only lowest-order sequential tunneling processes are taken into account, the model predicts that hysteretic switching might be observed if $V$ is kept sufficiently low that excited spin states are never populated (Fig. 4(b)). We investigated this experimentally at selected values of $V_g$ by applying a small constant bias $V$ and measuring the tunneling conductance while sweeping $B$, but still we observed no hysteresis. Possibly this difference is due to spin excitations from second- and higher-order cotunneling processes[14] that are neglected in the simulation.

In conclusion, we have measured electron tunneling in devices formed by inserting $Mn_{12}$ molecules into transistor structures. We find significant variations between devices, indicating that the sample fabrication process and the device environment may exert strong perturbations on the molecules. Nevertheless, the devices exhibit clear signatures of molecular magnetism in the tunneling spectra: zero-field splittings of energy levels and nonlinear variations of energies with magnetic field. We find that at most one excited magnetic level above the ground state is observed in the tunneling spectra. We do not observe magnetic hysteresis. All of these observations are in accord with a simple model of electron tunneling via a magnetic quantum system.

**Methods**

Our device fabrication is similar to techniques reported previously for making single-molecule transistors[1-3]. We first produced gate electrodes by depositing 40 nm of Al and oxidizing in air at room temperature. We used electron-beam lithography and liftoff to pattern Au wires 10 nm thick and 100 nm wide, and then exposed them to oxygen plasma for 30-120 s to remove any organic contaminants. Molecules were deposited on the samples by applying dilute (~100 $\mu$M) solutions of $Mn_{12}O_{12}(O_2CCH_3)_{16}(H_2O)_4$ ("$Mn_{12}Ac$") in acetonitrile or $Mn_{12}O_{12}(O_2CCHCl_2)_{16}(H_2O)_4$ ("$Mn_{12}Cl$") in methylene chloride for less than 1 minute. We then created molecular-sized gaps in the gold wires using electromigration[1] – by sweeping the applied voltage ($V$) at a rate of approximately 30 mV/s until the wire broke (typically at 0.7-0.8 V). Some samples underwent electromigration at a temperature $T = 1.5$ K in cryogenic vacuum, and others at room temperature, with magnetic properties observed in both cases. Detailed electrical transport measurements were performed at $T \leq 300$ mK. Magnetic fields were applied parallel to the plane of the substrate – see the inset of Fig. 1(b). The angle of the magnetic field with respect to the magnetic anisotropy axis of the molecule is expected to vary from device to device because the adsorption geometry of $Mn_{12}$ cannot be controlled during the deposition process.

**Acknowledgements**

The work at Harvard is supported by NSF, ARO, and the Packard Foundation. The work at Cornell is supported by the NSF through the Cornell Center for Materials Research and through use of the NNIN/Cornell Nanofabrication Facility and by ARO. The work at UC Berkeley is supported by NSF and Tyco Electronics.

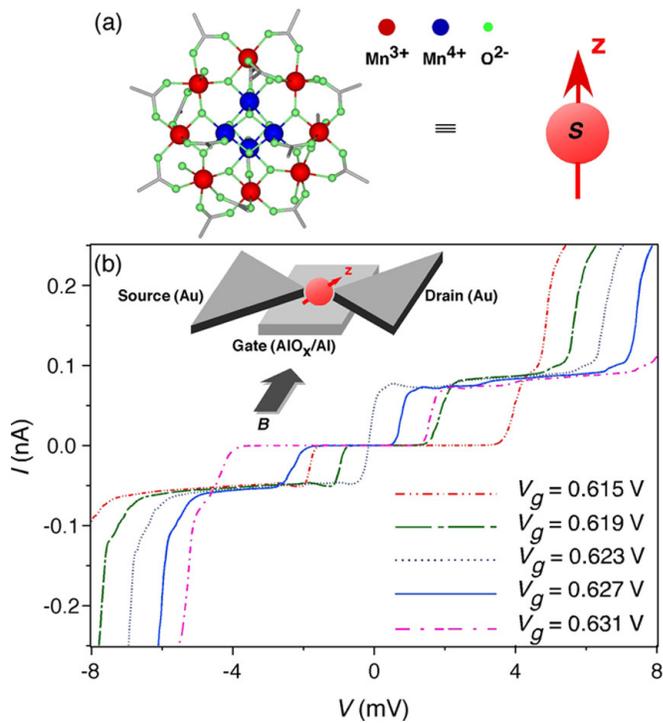

**Fig. 1 Mn$_{12}$ molecule and device characterization**. (a) Schematic diagram of a Mn$_{12}$ molecule. (b) Current-voltage (*I-V*) curves at selected values of gate voltage ($V_g$) for a Mn$_{12}$Ac transistor at 300 mK. (inset) Schematic of a single-molecule transistor.

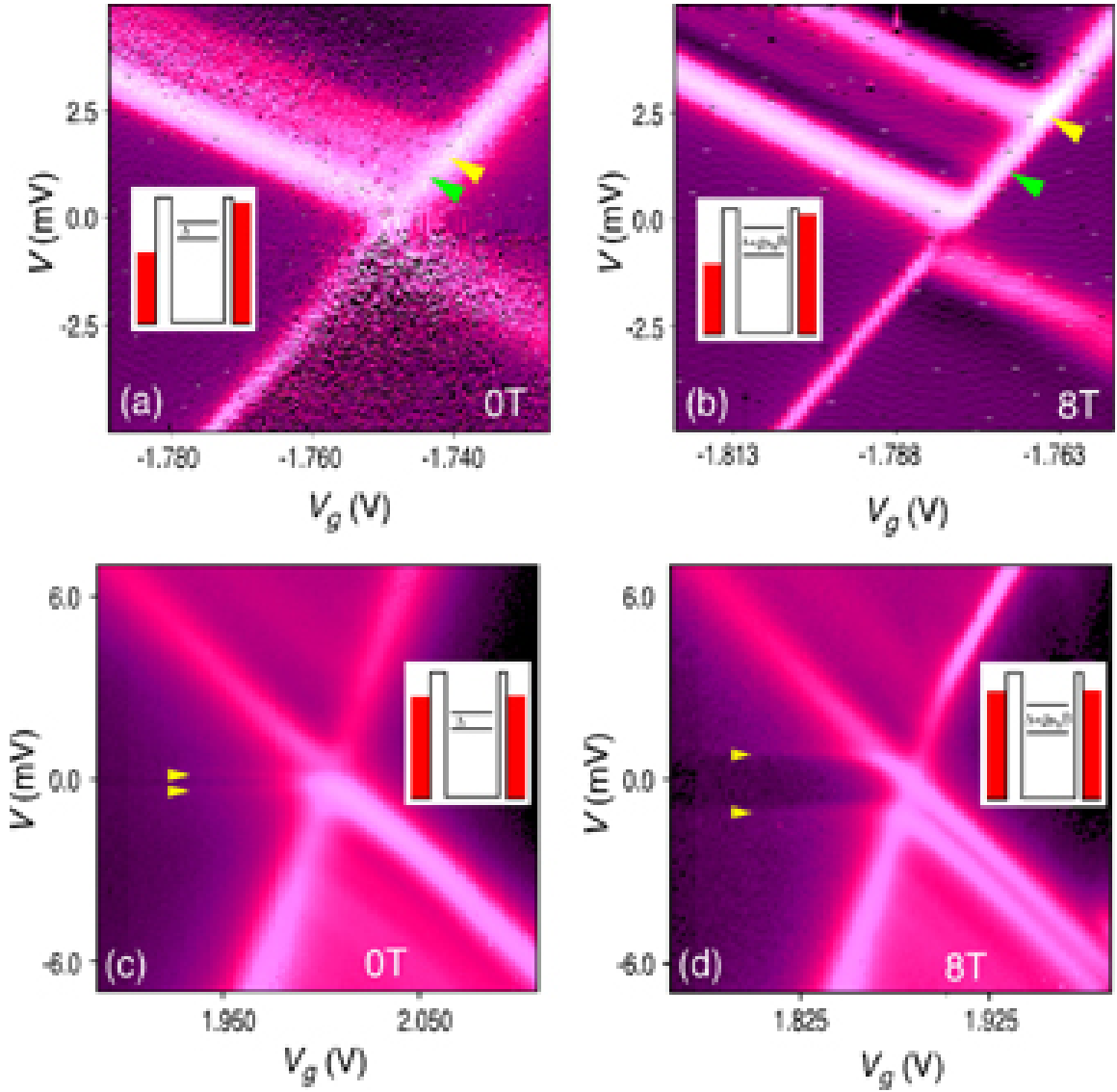

**Fig. 2 Differential conductance plots for Mn$_{12}$ transistors**. (a,b) $dI/dV$ vs. $V$ and $V_g$ for a Mn$_{12}$Ac transistor at $B = 0$ T and 8 T. Arrows (yellow and green) indicate excited energy states. The insets depict energy diagrams for the transport features. (c,d) $dI/dV$ vs. $V$ and $V_g$ for a Mn$_{12}$Cl transistor at $B = 0$ T and 8 T. Horizontal lines mark cotunneling features. The color scale in all panels varies from deep purple (10 nS) to light pink (200 nS). The scale in Fig. 2 (c,d) is logarithmic.

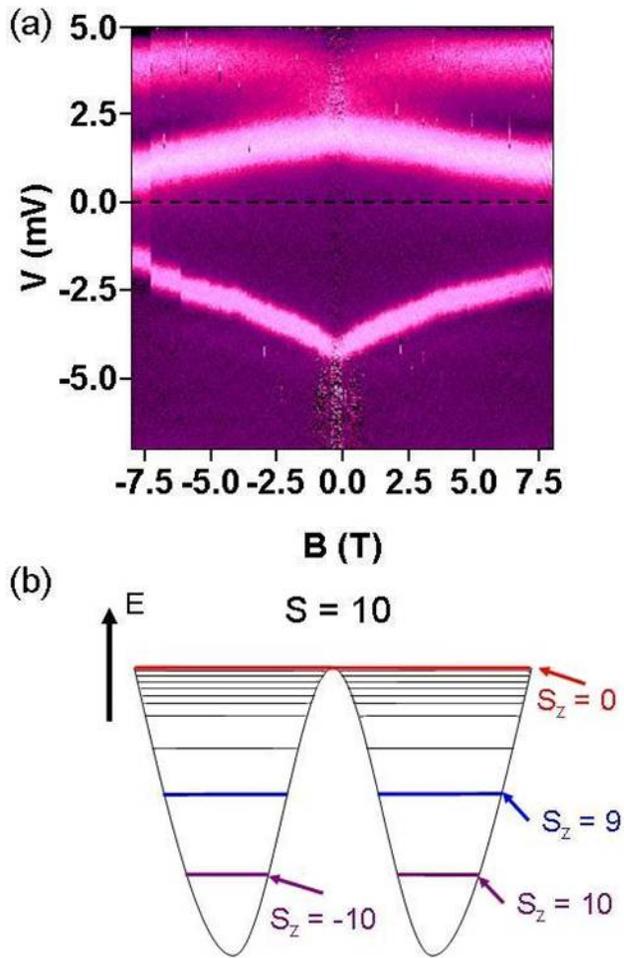

**Fig. 3 Energy-level spectrum as a function of magnetic field**. (a) Color plot of *dI/dV* vs. *V* and *B* at fixed $V_g$ = -1.768V for the same $Mn_{12}Ac$ device as in Fig. 2(a,b). The color scale varies from deep purple (10 nS) to light pink (200 nS) (b) Energy-level diagram for the $S$ = 10 multiplet in a $Mn_{12}$ molecule.

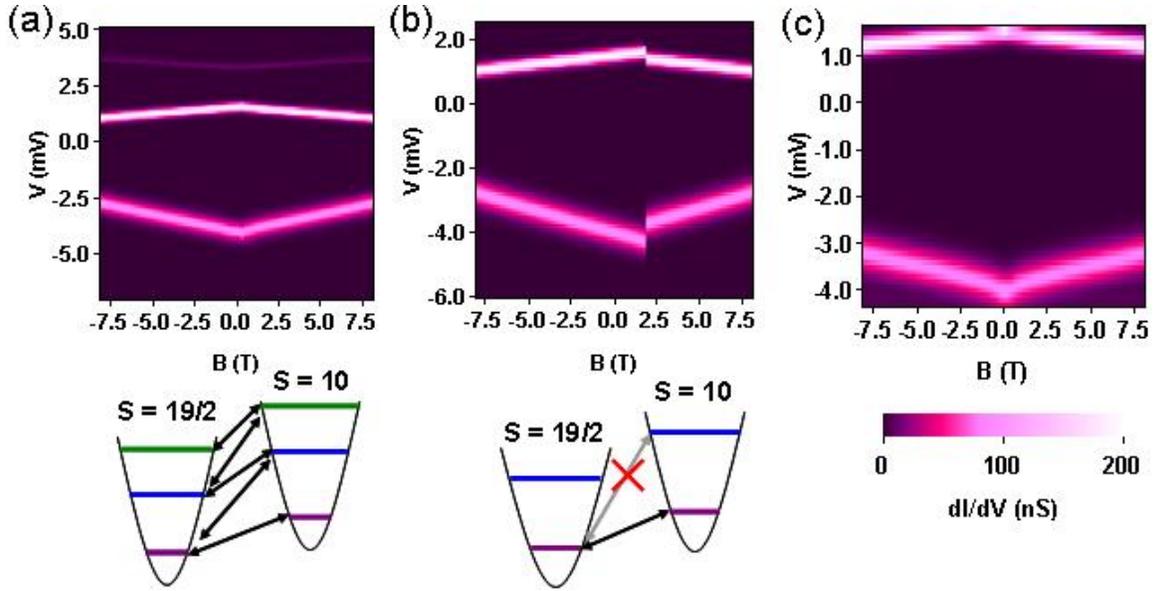

**Fig. 4 Results of numerical simulations**. (a, top) There is no hysteresis as a function of *B* when *V* is swept to values large enough to allow tunneling to excited magnetic states. (a, bottom) Diagram illustrating how energetically-allowed transitions provide a means to surmount the anisotropy barrier by climbing a ladder of spin states. (b) When the model includes only sequential tunneling processes (not cotunneling), a hysteretic jump in the level spectrum is predicted if *V* is kept sufficiently small that tunneling to excited spin states is not allowed. (c) Curvature in the dependence of energy levels on magnetic field occurs when the strength of magnetic anisotropy is different for the two charge states. Here $D_{N-1}$ = -0.071 meV and $D_N$ = -0.086 meV.